\documentstyle[aps,preprint]{revtex}
\begin{document}
\tightenlines
\title{Fresh inflation from five-dimensional vacuum state}
\author{Mauricio Bellini\footnote{E-mail address: mbellini@mdp.edu.ar;
bellini@ginette.ifm.umich.mx}}
\address{Instituto de F\'{\i}sica y Matem\'aticas, 
Universidad Michoacana de San Nicol\'as de Hidalgo,
AP:2-82, (58041) Morelia, Michoac\'an, M\'exico.}

\maketitle
\begin{abstract}
I study fresh inflation from a five-dimensional vacuum state, where the
fifth dimension is constant. In this framework, the universe can be seen
as inflating in a four-dimensional Friedmann-Robertson-Walker metric
embedding in a five-dimensional metric.
Finally, the
experimental
data $n_s=1$
(BOOMERANG-98 and MAXIMA-1, taken together COBE DMR),
are consistent with ${{\rm p}+\rho_t \over \rho_t}\simeq 1/3$
in the fresh inflationary
scenario.

\end{abstract}
\vskip 1cm
{\bf Introduction:}
In the last years higher-dimensional theories of gravity have received
much interest. For the most part, four-dimensional spacetime has been
extended
by the addition of several extra patial dimensions, usually taken to be
compact. On the other hand, Wesson and co-workers\cite{1} have given
new impetus to the study of five-dimensional gravity by adopting a different
approach, in which the extra dimension is not assumed to be compact.
The main question that they address is whether the four-dimensional
properties of matter can be viewed as being purely geometrical in origin.
This idea was originally introduced by Einstein\cite{2}.
Some years ago, Wesson and co-workers\cite{1}
suggested that the correct field equations
in five-dimensions are the vacuum Einstein field equations, and thus, the
higher-dimensionally stress-energy tensor is taken to be
identically zero.
In this approach, the five-dimensional
vacuum field equations, $G_{AB}=0$, give the four-dimensional field
equations $G_{\alpha\beta}=8\pi {\rm G} \  T_{\alpha\beta}$. So, the
four-dimensional properties of matter can be considered to be
purely geometrical in origin. 

In particular I consider a five-dimensional metric introduced by Ponce
de Leon\cite{3}
\begin{equation}\label{1}
ds^2=-\Psi^2 dt^2 + t^{2/\alpha} \Psi^{\frac{2}{1-\alpha}} dr^2 +
\left(\frac{\alpha}{\alpha -1}\right)^2 t^2 d\Psi^2,
\end{equation}
where $\Psi$ is a scalar, $\alpha$ is a constant and $dr^2 = dx^2+
dy^2+dz^2$. The interesting here is that the metric (\ref{1}) gives
$R_{AB}=0$. Here, the metric components are separable functions of $t$
and $\Psi$, and the global spatial curvature is considered as zero
(i.e., $k=0$). For this metric we obtain $G^{0}_{ \  0}
= 8\pi {\rm G} \  T^{0}_{ \  0}$ (with $T^{0}_{ \  0}=\rho_t$)
and $G^{1}_{ \  1} = 8 \pi {\rm G} \  T^{1}_{ \  1}$ (where
$T^{1}_{ \  1} = -{\rm p}$),
where the four-dimensional
metric is the intrinsic metric of the hypersurface $\Psi = {\rm const.}$
With these equations we obtain
$8\pi {\rm G}\rho_{t} = {3\over \alpha^2} {1\over \Psi^2 t^2}$ and
$8\pi {\rm G}{\rm p} = {2\alpha -3 \over \alpha^2} {1 \over \Psi^2 t^2}$. The pressure
${\rm p}$ and the total energy density $\rho_{t}$ are related by with the
equation of state
\begin{equation}
{\rm p} = \left(\frac{2\alpha - 3}{3}\right) \rho_{t}.
\end{equation}
I am interested in the case $0< \alpha < 3/2$, for which inflation can take
place due to the fact ${\rm p} <0$. In particular, I am interested in the study
of the fresh inflationary scenario\cite{4} for which the parameter
$\alpha$ is given by
\begin{equation}\label{3}
\alpha = - \frac{\dot H}{H^2} = \frac{3\dot\phi^2 + 4 \rho_r}{
2\rho_r + \dot\phi^2 + 2V(\phi)},
\end{equation}
where $\phi$ is the inflaton field, $\rho_r$ is the radiation energy
density, $V(\phi)$ is the scalar potential and the scale factor of the
universe is related to the Hubble parameter by $H= {\dot a \over a}$.
Hence, the scale factor evolves in time as $a \sim t^{1/\alpha}$, with
$0<\alpha < 1$. The fresh inflationary scenario was recently introduced
with the following characteristics\cite{4}.
Initially, the universe is not thermalized so that the radiation energy
density when fresh inflation starts is zero [$\rho_r(t=t_0)=0$]. We
understand the initial time to be the Planckian time ${\rm G}^{1/2}$. Later,
the universe will describe a second-order phase transition. Particle
production and heating ocurr together during the rapid expansion
of the universe. Hence, the radiation energy density grows during
fresh inflation ($\dot\rho_r >0$). The interaction between the inflaton
field and the particles produced during inflation provides slow-rolling
of the inflaton towards the minimum of the potential $V(\phi)$. Hence,
in the fresh inflationary model the slow-roll conditions are physically
well justified. The decay width ($\Gamma$) of the produced particles
grows with time, so when the inflaton approaches the minimum of the potential
there is no oscillation around the minimum energetic
configuration due to dissipation being too large with respect to the
Hubble parameter ($\Gamma \gg H$).
Hence, the thermal equilibrium holds for
$t \gg 10^7 \  {\rm G^{1/2}}$. In other words, in fresh inflation the
universe starts
from chaotic initial conditions and expands in an increasing
damped regime product of the interaction of the inflaton field with
other scalar field of a zero temperature initial state.
The interaction is represented in the four-dimensional
Lagrangian as ${\cal L}_{{\rm int}}
\sim - {\rm g}^2 \phi^2\varphi^2$, which describes
the interaction between the scalar field $\phi$ and the other
$\varphi$-scalar fields of a thermal bath.
The Lagrangian in fresh inflation is
\begin{equation}
{\cal L}= -\sqrt{-^{(4)}g} \left[ \frac{^{(4)}R}{16\pi {\rm G}} +
\frac{1}{2} g^{\mu\nu} \phi_{,\mu}\phi_{,\nu} + V(\phi) + {\cal
L}_{{\rm int}}\right],
\end{equation}
where $V(\phi) =[{\cal M}^2(0)/2]\phi^2+ [\lambda^2/4]\phi^4$, $^{(4)}R$ is
the four-dimensional scalar curvature, $^{(4)}g$ is the determinant of the
four-dimensional metric tensor,
${\rm G}=M^{-2}_p$ is the gravitational constant
and $M_p=1.2 \  10^{19} \  {\rm GeV}$
is the Planckian mass.
The inflaton field is really an effective field due to
$\phi=(\phi_i \phi_i)^{1/2}$. Furthermore, ${\cal M}^2(0)$ is
given by ${\cal M}^2_0$ plus renormalization counterterms in the
originary potential ${1\over 2}{\cal M}^2_0 (\phi_i\phi_i) + {\lambda^2
\over 4}(\phi_i\phi_i)^2$\cite{5}. The effective potential
is $V_{eff}(\phi) = [{\cal M}^2(\theta)/2]\phi^2+[\lambda^2/4]\phi^4$ (here,
$\theta$ is the temperature
and ${\cal M}^2(\theta) = {\cal M}^2(0)+ {(n+2) \over 12} \lambda^2
\theta^2$), such that $V_{eff}(\phi,\theta) =
V(\phi) + \rho_r(\theta,\phi)$.
The temperature increases
with the expansion of the universe because the inflaton transfers radiation
energy density to the bath with a rate larger than the expansion
of the universe.
So, the number of created particles $n$ [for
$\rho_r=(\pi^2/30) g_{\rm eff} \theta^4$], is given by
\begin{equation}
(n+2) = \frac{2\pi^2}{5\lambda^2} g_{{\rm eff}} \frac{\theta^2}{\phi^2},
\end{equation}
where $g_{{\rm eff}}$ denotes the effective degrees of freedom of the
prticles and it is assumed that $\varphi$ has no self-interaction.
We consider a Yukawa interaction $\delta =\dot\rho_r + 4H\rho_r=
\Gamma(\theta)\dot\phi^2$, where $\Gamma(\theta) ={g^4_{\rm eff} \over
192 \pi} \theta$\cite{6}.
But the crucial point here is that
this model attempts to build a bridge between the standard\cite{7} and
warm inflationary models\cite{8}, beginning from
chaotic initial conditions which provides naturality.

{\bf Fresh inflation from five-dimensional metric:}
Due to the fact that $p+\rho_t = {1\over 4\pi {\rm G}\alpha \tau^2} =
\dot\phi^2 + {4\over 3} \rho_r$ (where $\tau=\Psi t$ --- in the following
the overdot represents the derivative with respect to $\tau$),
we can use the expressions\cite{4}
\begin{eqnarray}
V(\phi)& =& \frac{9\left( 2-\alpha\right) \lambda^2}{32 \pi {\rm G}\alpha^2}
\phi^2 + \frac{\lambda^2}{4} \phi^4, \\
\rho_r & = & \left(\frac{\alpha}{2-\alpha}\right) V(\phi) -
\frac{3 \alpha^2}{2} \left(\frac{H^2}{H'}\right)^2 \frac{(3-\alpha)}{
(2-\alpha)},\\
H(\phi) & = & \frac{\lambda}{\alpha} \phi, \\
\phi(t) & = & \left(\lambda \tau \right)^{-1},
\end{eqnarray}
to obtain the value of the constant $\Psi$ in eq. (\ref{1})
\begin{equation}\label{psi}
\Psi^2 = \frac{2}{3}.
\end{equation}
Hence, taking the scale factor as $a(\tau) = \Psi^{{2(\alpha-1)\over
\alpha(1-\alpha)}} \tau^{1/\alpha}$
the five-dimensional metric in fresh inflation result
\begin{equation}
ds^2=-d\tau^2 + a^2(\tau) dr^2 + \left(\frac{\alpha}{\alpha-1}\right)^2
t^2 d\Psi^2,
\end{equation}
with $\Psi^2$ given by (\ref{psi}) and $0<\alpha<1$.
Hence, the effective four-dimensional metric being given by a
Friedmann-Robertson Walker (FRW) metric with a zero spatial curvature.

The fluctuations of the inflaton field $\delta\phi(\vec x,t)$ are
given by the equation of motion
\begin{equation}\label{e}
\ddot\delta\phi - \frac{1}{a^2} \nabla^2 \delta\phi + \left(3H
+ \Gamma\right)
\dot\delta\phi +  V''(\phi) \  \delta\phi =0.
\end{equation}
Here, the additional second term appears because the fluctuations
$\delta\phi$ are spatially inhomogeneous. The equation
for the modes
$\chi_k(\vec x,\tau) = \xi_k(\tau) e^{i\vec k.\vec x}$ and
$\chi^{*}_k(\vec x,\tau)=
e^{-i\vec k.\vec x} \xi^*_k(\tau)$,
of the redefined fluctuations
$\chi=a^{3/2} e^{{1\over 2}\int \Gamma d\tau} \  \delta \phi $
(which can be written as a Fourier expansion as)
\begin{equation}
\chi(\vec x,\tau) = \frac{1}{(2\pi)^{3/2}} {\Large\int} d^3k
\left[a_k \  \chi_k(\vec x,\tau)
+ a^{\dagger}_k \  \chi^*_k(\vec x,\tau)\right],
\end{equation}
is
\begin{equation}\label{xi}
\ddot\xi_k + \omega^2_k \xi_k=0,
\end{equation}
where $\omega^2_k = a^{-2}\left[k^2 - k^2_0\right]$ is the
squared frequency for each mode and
$k^2_0$ is given by 
\begin{equation}\label{k^2}
k^2_0(\tau) = a^2 \left[\frac{9}{4}\left(H+\Gamma/3\right)^2 +
3\left(\dot H+\dot\Gamma/3\right)- V''[\phi(t)]\right].
\end{equation}
Here,
the scale factor $a$ evolves as $a \sim \tau^p$, with $p=1/\alpha$ [see eq.
(\ref{3})], and
the time-dependent wave number $k_0(t)$ separates the infrared
(IR) and ultraviolet (UV) sectors.
Furthermore
($a_k$, $a^{\dagger}_k$) are respectively the annihilation and
creation operators. 
When inflation starts $\Gamma(\tau=\tau_0) \simeq 0$.
Furthermore the time derivative of the width decay $\dot\Gamma$
is nearly constant [$\dot\Gamma(\tau) \simeq
{\cal M}^2(0)$], so that, if we take
$\xi_k = \xi^{(0)}_k e^{\int g d\tau}$, the equation for $\xi^{(0)}_k$
can be approximated to
\begin{equation}\label{eq}
\ddot\xi^{(0)}_k +\left[\frac{k^2 \tau^{-2p}}{a^2_0 \tau^{-2p}_0}
- \left(\frac{9}{4}p^2 -3p -3\right) \tau^{-2}\right]\xi^{(0)}_k=0.
\end{equation}
The general solution (for $\nu \neq 0,1,2,...$)
for this equation is 
\begin{equation}\label{in}
\xi^{(0)}_k(\tau) = C_1 \sqrt{\frac{\tau}{\tau_0}}
{\cal H}^{(1)}_{\nu}\left[
\frac{k \tau^{1-p}}{a_0 \tau^{-p}_0(p-1)}\right]+ C_2
\sqrt{\frac{\tau}{\tau_0}} {\cal H}^{(2)}_{\nu}\left[
\frac{k \tau^{1-p}}{a_0 \tau^{-p}_0(p-1)}\right],
\end{equation}
where $\nu = {\sqrt{9p^2 -12p-11} \over 2(p-1)}$, which tends to
$3/2$ as $p \rightarrow \infty$ (i.e., $\nu \simeq 3/2$ for
$p \gg 1$), and $({\cal H}^{(1)}_{\nu}, {\cal H}^{(2)}_{\nu})$ are the
Hankel functions.
These functions take the small-argument
limit $\left.{\cal H}^{(2,1)}_{\nu}[x]\right|_{x\ll 1} \simeq
{(x/2)^{\nu} \over \Gamma(1+\nu)} \pm {i\over \pi}
\Gamma(\nu) \left(x/2\right)^{-\nu}$.
We can take the Bunch-Davis vacuum such that $C_1=0$ and
$C_2 = \sqrt{\pi/2}$\cite{BD}. 
Notice that $\xi^{(0)}_k$ is the solution for the
modes when the interaction is negligible ($\Gamma \propto \theta \simeq 0$).
The dependence of the Yukawa interaction is in the function $g(\tau)$, which
only takes into account the thermal effects.
The differential equation for $g$ is
\begin{equation}
g^2 + \dot g = \frac{3}{2} H \Gamma + \frac{\Gamma^2}{4},
\end{equation}
with initial condition $g(\tau=\tau_0)=0$.
The squared fluctuations for super Hubble scales ($k^2 \ll k^2_0$),
are given by
\begin{equation}\label{sf}
\left<\left(\delta\phi\right)^2\right> = \frac{a^{-3}}{2\pi^2}
F(\tau) {\Large\int}^{k_0(\tau)}_{0}
dk \  k^2 \xi^{(0)}_k \left(\xi^{(0)}_k\right)^*,
\end{equation}
where the asterisk denotes the complex conjugate, the function $F$ is
given by $F(\tau)=e^{-{\cal M}^2(0) \tau^2} e^{2\int g d\tau}$
(see Figure 1),
and
\begin{equation}
\xi^{(0)}_k \left(\xi^{(0)}_k\right)^*
\simeq \frac{2^{2\nu}}{\pi^2} \Gamma^2(\nu) \left[
\frac{a_o (p-1)}{\tau^p_0} \tau^{(p-1)} \right]^{2\nu} k^{-2\nu},
\end{equation}
so that the integral controlling the presence of infrared divergences is
${\Large\int}^{k_0(\tau)}_0 dk \  k^{2(1-\nu)}$ with
a power spectrum ${\cal P}_{<(\delta\phi)^2>} \sim k^{3-2\nu}$. Hence, 
the condition $n_s=3/2-\nu$ gives a spectral index $n_s \simeq 1$ according
with the experimental data\cite{prl} for $\nu \simeq 1/2$ (i.e., for
$p \simeq 2$).
This shows that for $1/2 \le \nu < 3/2$ (i.e., for $2 \le p < 3.3$), there
is no infrared divergence of the squared fluctuations (\ref{sf}).
However, for $\nu \ge 3/2$ (i.e., for $p\ge 3.3$), fresh inflation
predicts infrared divergence of the matter field fluctuations.
The figure shows $F(\tau)$ as a function of time for $p=2$ and
$p=3$. Notice that $F(\tau)$ increases as $p$ is more large.
The function $F(\tau)$ only depends on the temperature and
should be an analytical evidence
of superheavy particles produced during fresh inflation.\\

{\bf Final Comments:}
To summarize, 
the five-dimensional
vacuum field equations, $G_{AB}=0$,
are shown to induce four-dimensional fluid FRW field
equations $G_{\alpha\beta}=8\pi {\rm G} \  T_{\alpha\beta}$, so that the
four-dimensional properties of matter can be considered to be
purely geometrical in origin. 
In this framework, fresh inflation can be understood as a dissipative
inflating universe in a five-dimensional vacuum state where the
fifth dimension is constrained to $\Psi^2 = 2/3$.
The model studied here may provide the necessary number of $e$ folds
to explain the flatness/horizon problem for $\alpha \le 1/2$.
This inequality assures slow-roll
conditions during the fresh inflationary expansion of the universe. Notice
$\alpha \simeq 1/2$ (i.e., for $p\simeq 2$),
is the value which agrees with experimental
data $n_s \simeq 1$\cite{prl}. This value for $n_s$ is in good agreement
with ${{\rm p}+\rho_t \over \rho_t} \simeq 1/3=2\alpha/3$, which
is a confirmation that $\alpha$ is related with the properties
of matter\cite{*}.

\vskip 1cm
\noindent
Fig. 1: Temporal evolution of the function $F$ for $p=2$ (dashed
line)
and
$p=3$ (continuous line). The solution of $g$ was computed to
$g\gg {\dot\xi^{(0)}_k \over \xi^{(0)}_k}$.
All the calculations were made taking ${\cal M}^2(0)
=10^{-12} \  {\rm G}^{-1}$. 
\end{document}